\documentclass[12pt]{article}
\usepackage{amsfonts,amssymb,amsmath}
\def\dac{\displaystyle\frac}
\def\d{\partial}
\def\[{\left[}
\def\]{\right]}
\def\({\left(}
\def\){\right)}

\newcommand{\diag}{\mathop{\rm diag}\nolimits}

\begin{document}

\begin{center}
{\Huge\textbf{Anisotropic exponential cosmological solutions in$\vphantom{\dfrac12}$ 4th and 5th orders of Lovelock gravity}}\\
\textbf{I. V. Kirnos$\vphantom{\dfrac12}$}\\
\textit{Tomsk State University and V. E. Zuev Institute of Atmospheric Optics, Tomsk, Russia$\vphantom{\dfrac12}$}\\
kirnos@iao.ru, ikirnos@sibmail.com
\end{center}

\begin{abstract}
Anisotropic exponential cosmological solutions for a space of arbitrary dimension filled with ordinary matter in the 4th and 5th orders of Lovelock gravity are obtained. Also we have supposed a generalization of such solutions on an arbitrary order. All the solutions are represented as a set of conditions on Hubble parameters.
\end{abstract}

\section*{Introduction}
A huge amount of investigations in cosmology during the last quarter of a sentury motivate theoreticians to develop new gravitational theories. Many of these theories are modifications of General Relativity (GR).

Any modification of GR includes additional fields (such as scalar field, torsion, second metric tensor etc.) or higher derivatives in field equations or extra spatial dimensions. It is impossible to avoid all above-mentioned features.

Lovelock gravity \cite{lovelock} has no additional fieds and no higher derivatives. It is based on\\
\textbf{Lovelock theorem}\\
If in $n$-dimensional riemannian space one needs tensor $G_{\mu\nu}$ (gravity field tensor) with the following features:
\begin{enumerate}
\item $G_{\mu\nu}$ is symmetric: $G_{\mu\nu}=G_{\nu\mu}$,
\item $G_{\mu\nu}$ is divergence free: $\nabla^\mu G_{\mu\nu}=0$,
\item  $G_{\mu\nu}$ is a concomitant of the metric tensor and its first two derivatives:\\  $G_{\mu\nu}= G_{\mu\nu}(g_{\mu\nu},\d_\alpha g_{\mu\nu},\d_\alpha\d_\beta g_{\mu\nu})$,
\end{enumerate}
then general expression for $G_{\mu\nu}$ is
\begin{equation}\label{lovelock}
G^\mu_{\phantom{\mu}\nu}=\sum_{p=1}^{m-1}\alpha_p
G^{(p)\mu}_{\phantom{(p)\mu}\nu}+\Lambda
\delta^{\mu}_{\phantom{\mu}\nu},\end{equation}
where
$$ m=\frac1 2 n,\qquad\mbox{if $n$ is even,}$$ $$ m=\frac1 2
(n+1),\qquad\mbox{if $n$ is odd,}$$
\begin{equation}\label{lovelocktensor} G^{(p)\mu}_{\phantom{(p)\mu}\nu}=\delta^{\mu\lambda_1\lambda_2
\cdots\lambda_{2p}}_{\nu\sigma_1\sigma_2\cdots\sigma_{2p}}R_
{\lambda_1\lambda_2}^{\phantom{\lambda_1\lambda_2}\sigma_1\sigma_2}R_{\lambda_3\lambda_4}^
{\phantom{\lambda_3\lambda_4}\sigma_3\sigma_4}\cdots
R_{\lambda_{2p-1}\lambda_{2p}}^{\phantom{\lambda_{2p-1}\lambda_{2p}}\sigma_{2p-1}\sigma_{2p}},\end{equation}
 $\alpha_p$, $\Lambda$ are arbitrary constants,
$\delta^{\mu_1\cdots\mu_k}_{\nu_1\cdots\nu_k}$ is multidimensional delta-symbol, which equals to one, if
$\nu_1\cdots\nu_k$ is even permutation of $\mu_1\cdots\mu_k$, equals to minus one, if odd, and equals to zero in other cases. We will call tensor (\ref{lovelocktensor}) a $p$-th order Lovelock tensor.

It is easy to understand that in 4-dimensional spacetime only 0-th and 1-st Lovelock tensors are nonzero, so
\begin{equation}
G_{\mu\nu}=\alpha(R_{\mu\nu}-\dfrac12Rg_{\mu\nu})+\beta g_{\mu\nu}
\end{equation}
and we have ordinary Hilbert-Einstein equations with cosmological term.

Hence if we need new results in Lovelock gravity then we should consider spacetimes with 5 dimensions or more. But in such a case we should explain an invisibility of extra dimensions. We can do this by means of Kluza-Klein approach: extra spacial dimensions are considered as closed and small.

But such an approach means that space is anisotropic. So we should look for anisotropic solutions of gravity field equations. And it is interesting to consider maximally anisotropic space: it might arise isotropization of 3-dimensional visible space or invisible dimensions might behave in different ways.
\section{Earlier-obtained anisotropic cosmological solutions}
\subsection{Power-law solutions}

%Earlier some anisotropic cosmological solutions in Lovelock gravity have already obtained. 
Consider metric tensor with power-law scale factors:
\begin{equation}g_{\mu\nu}=\diag\{-1,t^{2p_1},\ldots,t^{2p_n}\},\end{equation}
where $p_i,$ $i=1,\ldots,n$ are constant values (power-law parameters). In the first order (i. e. in ordinary GR) we have Kasner solution for an empty space \cite{kasner}:
\begin{equation}\label{Kasner-1}\sum_ip_i=1,\quad\sum_{i<j}p_ip_j=0.\end{equation}
and Jacobs solution for maximally stiff fluid ($p=w\rho$) \cite{jacobs}:
\begin{equation}\label{Jacobs-1}w=1,\quad\sum_ip_i=1,\quad
\sum_{i<j}p_ip_j=-\dac{1}{4\alpha_1}\cdot\dac{8\pi G}{c^4}\varepsilon_0,\end{equation}
where $\varepsilon_0$ is initial matter density.
In the second order equations $\alpha_2{G^{(2)}}_{\mu\nu}=\varkappa T_{\mu\nu}$ have an analogue of Kasner solution
\begin{equation}\label{Kasner-2}\sum_ip_i=3,\quad\sum_{i<j<k<l}p_ip_jp_kp_l=0,\end{equation}
discovered by N. Deruelle \cite{deruelle} and rediscovered by A. Toporensky and P. Tretyakov \cite{toporensky-tretyakov} and an analogue of Jacobs solution
\begin{equation}\label{Jacobs-2}w=1/3,\quad\sum_ip_i=3,\quad
\sum_{i<j<k<l}p_ip_jp_kp_l=-\dac{1}{96\alpha_2}\cdot\dac{8\pi
G}{c^4}\varepsilon_0,\end{equation}
discovered by author \cite{kirnos-makarenko-pavluchenko-toporensky}.

For an arbitrary $p$-th order ($\alpha_p{G^{(p)}}_{\mu\nu}=\varkappa T_{\mu\nu}$, hereinafter $\varkappa\equiv8\pi G/c^4$) Kasner solution has been generalized by S. Pavluchenko \cite{pavluchenko} and, independently, by author \cite{arbitrary_order}:
\begin{equation}\label{Kasner}\sum_i p_i=2p-1,\quad\sum_{i_1<i_2<\ldots<i_{2p}}p_{i_1}p_{i_2}\cdots p_{i_{2p}}=0,\end{equation} 
Jacobs solution has been generalized by author \cite{arbitrary_order}:
\begin{equation}\label{Jacobs}\sum_i p_i=2p-1,\quad\sum_{i_1<i_2<\ldots<i_{2p}}p_{i_1}p_{i_2}\cdots p_{i_{2p}}=-\dac{1}{\alpha_p 2^p(2p)!}\cdot\dac{8\pi G}{c^4}\varepsilon_0.\end{equation}

Unfortunately all these solutions involve only one order of Lovelock gravity (without involving lower orders) and, secondly, solutions with matter order specific value of EoS parameter $w$.

\subsection{Exponential solutions}

Such shortcomings are absent for exponential solutions:
\begin{equation}g_{\mu\nu}=\diag\{-1,e^{2H_1t},\ldots,e^{2H_nt}\},\end{equation}
where $H_i,$ $i=1,\ldots,n$ are constant values (Hubble parameters). In the second order equations
\begin{equation}{G^{(1)}}_{\mu\nu}+\alpha_2{G^{(2)}}_{\mu\nu}=\varkappa T_{\mu\nu}\end{equation}
have solution \cite{kirnos-pavluchenko-toporensky}
\begin{equation}\begin{array}{l}
\sum_i H_i=0,\\
\sum_i H_i^2=(1-3w)\varkappa_0,\vphantom{\dfrac12}\\
\sum_i H_i^4=\dfrac{w-1}{2\alpha_2}\varkappa_0+\dfrac{(1-3w)^2}{2}\varkappa_0^2
\end{array}\end{equation}
(hereinafter $\varkappa_0\equiv\dfrac{8\pi G}{c^4}\varepsilon_0$).
In the third order equations
\begin{equation}\alpha_1{G^{(1)}}_{\mu\nu}+\alpha_2{G^{(2)}}_{\mu\nu}+\alpha_3{G^{(3)}}_{\mu\nu}=\varkappa T_{\mu\nu}\end{equation}
have solution \cite{arbitrary_order}
\begin{equation}\label{3_order}\begin{array}{l}
\sigma_1=0,\\
\sigma_4=\dfrac12\sigma_2^2-\dfrac{\alpha_1}{2\alpha_2}\sigma_2-\dfrac{1-5w}{16\alpha_2}\varkappa_0,\vphantom{\dfrac{\dfrac12}{\dfrac12}}\\
\sigma_6=\dfrac13\sigma_3^2+\dfrac14\sigma_2^3-\dfrac{3\alpha_1}{8\alpha_2}\sigma_2^2+\dfrac{1}{96}\(\dfrac{\alpha_1}{\alpha_2}-\dfrac{9(1-5w)}{2\alpha_2}\varkappa_0\) \sigma_2+\dfrac{1-3w}{384\alpha_3}\varkappa_0
\end{array}\end{equation}
(hereinafter $\sigma_s\equiv\sum_i H_i^s,$ $s=1,2,\ldots$).

\section{New solutions in 4th and 5th orders}

In this paper we will obtain exponential solutions in 4th and 5th orders of Lovelock gravity. Firstly define the notations:
\begin{equation}\sigma_s\equiv\sum_i H_i^s,\end{equation}
\begin{equation}\zeta_p\equiv\sum_{i_1}H_{i_1}\sum_{i_2\neq i_1}H_{i_2}\sum_{i_3\neq i_1,i_2}H_{i_3}\cdots\sum_{i_p\neq i_1,i_2,\ldots i_{p-1}}H_{i_p},\end{equation}
\begin{equation}\zeta_{p(j)}\equiv\sum_{i_1\neq j}H_{i_1}\sum_{i_2\neq j,i_1}H_{i_2}\sum_{i_3\neq j,i_1,i_2}H_{i_3}\cdots\sum_{i_p\neq j,i_1,i_2,\ldots i_{p-1}}H_{i_p},\end{equation}
\begin{equation}\eta_{p(j)}\equiv\sum_{i_1\neq j}H_{i_1}^2\sum_{i_2\neq j,i_1}H_{i_2}\sum_{i_3\neq j,i_1,i_2}H_{i_3}\cdots\sum_{i_p\neq j,i_1,i_2,\ldots i_{p-1}}H_{i_p},\end{equation}
where $s$, $p$ are arbitrary integers, the summation is over all the spatial dimensions. And find a relation between these values:
\begin{equation}\label{zeta-recursion}\begin{array}{l}\displaystyle\zeta_p=\sum_{i_1}H_{i_1}\cdots\sum_{i_{p-1}\neq\ldots}H_{i_{p-1}}(\sigma_1-H_{i_1}-H_{i_2}-\cdots-H_{i_{p-1}})=\\ \displaystyle\quad{}=\sigma_1\zeta_{p-1}-(p-1)\sum_{i_1}H_{i_1}\cdots\sum_{i_{p-2}\neq\ldots}H_{i_{p-2}} \sum_{i_{p-1}\neq\ldots}H_{i_{p-1}}^2=\\ \displaystyle\quad{}=\sigma_1\zeta_{p-1}-(p-1)\sum_{i_1}H_{i_1}\cdots\sum_{i_{p-2}\neq\ldots}H_{i_{p-2}}(\sigma_2-H_{i_1}^2-\cdots-H_{i_{p-2}})=\\ \displaystyle\quad{}=\sigma_1\zeta_{p-1}-(p-1)\sigma_2\zeta_{p-2}+(p-1)(p-2)\sum_{i_1}H_{i_1}\cdots\sum_{i_{p-3}\neq\ldots}H_{i_{p-3}}\sum_{i_{p-2}\neq\ldots}H_{i_{p-2}}^3=\\ \displaystyle\quad{}=\cdots=
\sum_{k=1}^{p-2}(-1)^{k-1}\dac{(p-1)!}{(p-k)!}\sigma_k\zeta_{p-k}+(-1)^{p-2}(p-1)(p-2)\cdots 4\cdot 3\cdot 2\times\\ \displaystyle\quad{}\times\sum_{i_1}H_{i_1}(\sigma_{p-1}-H_{i_1}^{p-1})=\sum_{k=1}^p(-1)^{k-1}\dac{(p-1)!}{(p-k)!}\sigma_k\zeta_{p-k},\end{array}\end{equation}
where it is noted: $\zeta_0\equiv 1$. Thus,
\begin{equation}\begin{array}{l}\zeta_0\equiv 1,\\
\displaystyle\zeta_p=\sum_{k=1}^p(-1)^{k-1}\dac{(p-1)!}{(p-k)!}\sigma_k\zeta_{p-k}.\end{array}\end{equation}
Now express $\zeta_p$ (we use condition $\sigma_1=0$ obtained in \cite{kirnos-pavluchenko-toporensky}):
\begin{equation}\label{zeta_p}\begin{array}{l}\zeta_0=1,\qquad\zeta_1=0,\qquad \zeta_2=-\sigma_2,\qquad\zeta_3=2\sigma_3,\qquad\zeta_4=3\sigma_2^2-6\sigma_4,\\ \zeta_5=-20\sigma_3\sigma_2+24\sigma_5,\qquad \zeta_6=-15\sigma_2^3+90\sigma_4\sigma_2+40\sigma_3^2-120\sigma_6,\vphantom{\dfrac12}\\
\zeta_7=210\sigma_2^2\sigma_3-504\sigma_2\sigma_5-420\sigma_3\sigma_4+720\sigma_7,\\
\zeta_8=105\sigma_2^4-1260\sigma_4\sigma_2^2-1120\sigma_3^2\sigma_2+3360\sigma_6\sigma_2+
2688\sigma_5\sigma_3+1260\sigma_4^2-5040\sigma_8,\vphantom{\dfrac12}\\
\zeta_9=-2520\sigma_3\sigma_2^3+9072\sigma_5\sigma_2^2+12096\sigma_4\sigma_3\sigma_2-25920\sigma_7\sigma_2+2240\sigma_3^3+20160\sigma_6\sigma_3-\\
\qquad\quad{}\vphantom{\dfrac12}-18144\sigma_5\sigma_4+40320\sigma_9,\\
\zeta_{10}=-945\sigma_2^5+18900\sigma_4\sigma_2^3+25200\sigma_3^2\sigma_2^2-75600\sigma_6\sigma_2^2-120960\sigma_5\sigma_3\sigma_2-
56700\sigma_4^2\sigma_2 +\\ \qquad\quad{}+\vphantom{\dfrac12} 226800\sigma_8\sigma_2-50400\sigma_4\sigma_3^2+172800\sigma_7\sigma_3+151200\sigma_6\sigma_4+72576\sigma_5^2-362880\sigma_{10}.
\end{array}\end{equation}

Acting as in (\ref{zeta-recursion}) we obtain
\begin{equation}
\displaystyle\zeta_{p(j)}=\sum_{k=1}^p(-1)^{k-1}\dac{(p-1)!}{(p-k)!}(\sigma_k-H_j^k)\zeta_{p-k(j)},\end{equation}
\begin{equation}
\displaystyle\eta_{p(j)}=\sum_{k=1}^p(-1)^{k-1}\dac{(p-1)!}{(p-k)!}(\sigma_{k+1}-H_j^{k+1})\zeta_{p-k(j)}.\end{equation}
Using these equations one can obtain
\begin{equation}\label{zeta_p(j)}\begin{array}{l}\zeta_{1(j)}=-H_j,\qquad \zeta_{2(j)}=2H_j^2-\sigma_2,\qquad \zeta_{3(j)}=-6H_j^3+3\sigma_2H_j+2\sigma_3,\vphantom{\dac 1 2}\\ \zeta_{4(j)}=24H_j^4-12\sigma_2H_j^2-8\sigma_3H_j+3\sigma_2^2-6\sigma_4,\vphantom{\dac 1 2}\\ \zeta_{5(j)}=-120H_j^5+60\sigma_2H_j^3+40\sigma_3H_j^2-15\sigma_2^2H_j+ 30\sigma_4H_j-20\sigma_3\sigma_2+24\sigma_5,\vphantom{\dac 1 2}\\
\zeta_{6(j)}=720H_j^6-360\sigma_2H_j^4-240\sigma_3H_j^3+90\sigma_2^2H_j^2-180\sigma_4H_j^2+
120\sigma_3\sigma_2H_j-\vphantom{\dac 1 2}\\ \qquad\quad{}-144\sigma_5H_j-15\sigma_2^3+90\sigma_4\sigma_2+40\sigma_3^2-120\sigma_6,\vphantom{\dac 1 2}\\
\zeta_{7(j)}=-5040H_j^7+2520\sigma_2H_j^5+1680\sigma_3H_j^4- 630\sigma_2^2H_j^3+1260\sigma_4H_j^3-840\sigma_3\sigma_2H_j^2+\vphantom{\dac 1 2}\\ \qquad\quad{}+1008\sigma_5H_j^2+105\sigma_2^3H_j-630\sigma_4\sigma_2H_j-280\sigma_3^2H_j+
840\sigma_6H_j+210\sigma_3\sigma_2^2-\vphantom{\dac 1 2}\\ \qquad\quad{}-504\sigma_5\sigma_2-420\sigma_4\sigma_3+720\sigma_7,\vphantom{\dac 1 2}\\
\zeta_{8(j)}=40320H_j^8-20160\sigma_2H_j^6-13440\sigma_3H_j^5+5040\sigma_2^2H_j^4-10080\sigma_4H_j^4+6720\sigma_3\sigma_2H_j^3-\vphantom{\dac 1 2}\\ \qquad\quad{}\vphantom{\dfrac12}-8064\sigma_5H_j^3-840\sigma_2^3H_j^2+5040\sigma_4\sigma_2H_j^2+2240\sigma_3^2H_j^2-6720\sigma_6H_j^2- 1680\sigma_3\sigma_2^2H_j+\vphantom{\dac 1 2}\\ \qquad\quad{}+4032\sigma_5\sigma_2H_j+3360\sigma_4\sigma_3H_j-5760\sigma_7H_j+105\sigma_2^4-1260\sigma_4\sigma_2^2-1120\sigma_3^2\sigma_2+\vphantom{\dac 1 2}\\ \qquad\quad{}+3360\sigma_6\sigma_2+ 2688\sigma_5\sigma_3+ 1260\sigma_4^2-5040\sigma_8,\vphantom{\dfrac12}\vphantom{\dac 1 2}\\
\displaystyle\zeta_{9(j)}=\sum_{s_1l_1+\ldots+s_kl_k+m=9}(-1)^{l_1+\ldots+l_k+9}\dfrac{9!}{s_1^{l_1}\cdot\ldots\cdot s_k^{l_k}l_1!\cdot\ldots\cdot l_k!}\sigma_{s_1}^{l_1}\cdot\ldots \cdot\sigma_{s_k}^{l_k}H_j^m,\vphantom{\dac 1 2}
\end{array}\end{equation}
\begin{equation}\label{eta_p(j)}\begin{array}{l}
\eta_{3(j)}=3\sigma_2H_j^2-\sigma_2^2-6H_j^4+2\sigma_3H_j+3\sigma_4,\vphantom{\dac 1 2}\\
\eta_{5(j)}=-120H_j^6+60\sigma_2H_j^4-15\sigma_2^2H_j^2-20\sigma_3\sigma_2H_j+ 40\sigma_3H_j^3+30\sigma_4H_j^2+\vphantom{\dac 1 2}\\ \qquad\qquad{}+3\sigma_2^3-18\sigma_4\sigma_2-8\sigma_3^2+24\sigma_5H_j+24\sigma_6,\vphantom{\dac 1 2}\\
\eta_{7(j)}=2520\sigma_2H_j^6-630\sigma_2^2H_j^4-840\sigma_3\sigma_2H_j^3+105\sigma_2^3H_j^2-
630\sigma_4\sigma_2H_j^2+\vphantom{\dac 1 2}\\ \qquad\quad{}+210\sigma_3\sigma_2^2H_j-504\sigma_5\sigma_2H_j-15\sigma_2^4+
180\sigma_4\sigma_2^2+160\sigma_3^2\sigma_2-\vphantom{\dac 1 2}\\ \qquad\quad{}-480\sigma_6\sigma_2-5040H_j^8+
1680\sigma_3H_j^5+1260\sigma_4H_j^4+1008\sigma_5H_j^3-280\sigma_3^2H_j^2+\vphantom{\dac 1 2}\\ \qquad\quad{}+
840\sigma_6H_j^2-420\sigma_4\sigma_3H_j-384\sigma_5\sigma_3-180\sigma_4^2+
720\sigma_7H_j+720\sigma_8,\vphantom{\dac 1 2}\\
\eta_{9(j)}=H_j\zeta_{9(j)}+105\sigma_2^5-2100\sigma_4\sigma_2^3-2800\sigma_3^2\sigma_2^2+8\cdot7\cdot6\cdot5\cdot5\sigma_6\sigma_2^2+ 10\cdot8\cdot7\cdot6\cdot4\sigma_5\sigma_3\sigma_2+\vphantom{\dac 1 2}\\ \qquad\quad{}+10\cdot9\cdot7\cdot10\sigma_4^2\sigma_2-7!\cdot5\sigma_8\sigma_2+10\cdot8\cdot7\cdot5\cdot2\sigma_4\sigma_3^2- 10\cdot8\cdot6\cdot5\cdot4\cdot2\sigma_7\sigma_3-\vphantom{\dfrac12}\\ \qquad\quad{}-10\cdot8\cdot7\cdot6\cdot5\sigma_6\sigma_4+8!\sigma_{10}-8\cdot7\cdot6\cdot4\cdot3\cdot2\sigma_5^2,
\end{array}\end{equation}

Now consider field equations. Taking metrics in the form
\begin{equation}g_{\mu\nu}=\diag\{-1,e^{2H_1t},\ldots,e^{2H_nt}\},\end{equation}
we have
\begin{equation}\label{G00-exp-3}
{G^{(p)0}}_0=2^p(2p)!\sum_{i_1<i_2<\cdots<i_{2p}}H_{i_1}H_{i_2}\cdots H_{i_{2p}}=2^p\zeta_{2p},\end{equation}
\begin{equation}\begin{array}{ll}\displaystyle{G^{(p)j}}_j&\displaystyle=2^p(2p)!\sum_{\substack{i_2,i_3,\ldots,i_{2p}\neq j\\ i_2<i_3<\cdots<i_{2p}}} H_{i_2}^2H_{i_3}H_{i_4}\cdots H_{i_{2p}}+\\ &\displaystyle\quad{}+2^p(2p)!\sum_{\substack{i_1,i_2,\ldots,i_{2p}\neq j\\ i_1<i_2<\cdots<i_{2p}}} H_{i_1}H_{i_2}\cdots H_{i_{2p}}=2^p\cdot 2p\eta_{2p-1(j)}+2^p\zeta_{2p(j)}=\vphantom{\dac 1 2}\\ &\displaystyle\quad{}=2^p\eta_{2p-1(j)}-2^pH_j\zeta_{2p-1(j)}.\vphantom{\dac 1 2}\end{array}\end{equation}
Using (\ref{zeta_p}), (\ref{zeta_p(j)}), (\ref{eta_p(j)}), one can obtain
\begin{equation}\begin{array}{l}{G^{(1)j}}_j=2\sigma_2,\qquad {G^{(2)j}}_j=8\sigma_4-4\sigma_2^2,\\ {G^{(3)j}}_j=24\sigma_2^3-144\sigma_4\sigma_2-64\sigma_3^2+192\sigma_6,\vphantom{\dac 1 2}\\
{G^{(4)j}}_j=16(-15\sigma_2^4+180\sigma_4\sigma_2^2+160\sigma_3^2\sigma_2-480\sigma_6\sigma_2-
384\sigma_5\sigma_3-180\sigma_4^2+720\sigma_8),\vphantom{\dac 1 2}\\
{G^{(5)j}}_j=32\[\vphantom{\dfrac12}105\sigma_2^5-2100\sigma_4\sigma_2^3-2800\sigma_3^2\sigma_2^2+8400\sigma_6\sigma_2^2+ 560\cdot24\sigma_5\sigma_3\sigma_2+ 6300\sigma_4^2\sigma_2-\right.\vphantom{\dac 1 2}\\ \left.\qquad\quad{}-5\cdot7!\sigma_8\sigma_2+5600\sigma_4\sigma_3^2-48\cdot400\sigma_7\sigma_3-56\cdot300\sigma_6\sigma_4+8!\sigma_{10}-\dfrac{8!}{5}\sigma_5^2\].\end{array}\end{equation}

Thus equations
\begin{equation}\alpha_1{G^{(1)\mu}}_\nu+\alpha_2{G^{(2)\mu}}_\nu+\alpha_3{G^{(3)\mu}}_\nu+\alpha_4{G^{(4)\mu}}_\nu+\alpha_5{G^{(5)\mu}}_\nu=\varkappa {T^\mu}_\nu\end{equation}
take form
\begin{equation}\label{xi-eq}\left\{\begin{array}{l}-\xi_1-3\xi_2-5\xi_3-7\xi_4-9\xi_5=-\varkappa\varepsilon_0,\\
\xi_1+\xi_2+\xi_3+\xi_4+\xi_5=w\varkappa\varepsilon_0,\end{array}\right.\end{equation}
where
\begin{equation}\begin{array}{l}
\xi_1\equiv2\alpha_1\sigma_2,\qquad\xi_2\equiv-4\alpha_2(\sigma_2^2-2\sigma_4),\\ \xi_3\equiv-8\alpha_3(-3\sigma_2^3+18\sigma_4\sigma_2+8\sigma_3^2-24\sigma_6),\vphantom{\dfrac12}\\
\xi_4\equiv16\alpha_4(-15\sigma_2^4+180\sigma_4\sigma_2^2+160\sigma_3^2\sigma_2-480\sigma_6\sigma_2-
384\sigma_5\sigma_3-180\sigma_4^2+720\sigma_8),\vphantom{\dfrac12}\\
\xi_5\equiv32\alpha_5\[105\sigma_2^5-2100\sigma_4\sigma_2^3-2800\sigma_3^2\sigma_2^2+8400\sigma_6\sigma_2^2+ 13440\sigma_5\sigma_3\sigma_2+6300\sigma_4^2\sigma_2 -\right.\vphantom{\dfrac12}\\ \left.\qquad\quad{}- 25200\sigma_8\sigma_2+5600\sigma_4\sigma_3^2-19200\sigma_7\sigma_3-16800\sigma_6\sigma_4-8064\sigma_5^2+40320\sigma_{10}\].\vphantom{\dfrac12}
\end{array}\end{equation}

In the 3rd order of Lovelock gravity ($\alpha_4=\alpha_5=0$) it is easy to obtain
\begin{equation}\xi_2=-2\xi_1+\dac{5w-1}{2}\varkappa\varepsilon_0,\qquad \xi_3=\xi_1+\dac{1-3w}{2}\varkappa\varepsilon_0.\end{equation}
From these equations we have solution (\ref{3_order}).

In the 4th order ($\alpha_5=0$) one can obtain
\begin{equation}\begin{array}{l}\xi_3=-2\xi_2-3\xi_1+\dfrac{7w-1}{2}\varkappa\varepsilon_0,\\
\xi_4=\xi_2+2\xi_1+\dfrac{1-5w}{2}\varkappa\varepsilon_0,\end{array}\end{equation}
so solution is
\begin{equation}\begin{array}{l}\sigma_1=0,\vphantom{\dfrac12}\\
\sigma_6=-\dfrac 1 8 \sigma_2^3+\dfrac34\sigma_4\sigma_2+\dfrac13\sigma_3^2+\dfrac{\alpha_2}{24\alpha_3}\(\sigma_2^2-2\sigma_4\)-\dfrac{\alpha_1}{32\alpha_3}\sigma_2-\dfrac{1-7w}{384\alpha_3}\varkappa\varepsilon_0,\\
\sigma_8=-\dfrac{1}{16}\sigma_2^4+\dfrac14\sigma_4\sigma_2^2+\dfrac{\alpha_2}{36\alpha_3}\(\sigma_2^3-2\sigma_4\sigma_2\)-\dfrac{\alpha_1}{48\alpha_3}\sigma_2^2- \dfrac{1-7w}{576\alpha_3}\sigma_2\varkappa\varepsilon_0+\dfrac{8}{15}\sigma_5\sigma_3+\vphantom{\dfrac{\dfrac12}{\dfrac12}}\\
\qquad\quad{}+\dfrac14\sigma_4^2-\dfrac{\alpha_2}{2880\alpha_4}\(\sigma_2^2-2\sigma_4\)+ \dfrac{\alpha_1}{2880\alpha_4}\sigma_2+\dfrac{1-5w}{23040\alpha_4}\varkappa\varepsilon_0.\end{array}\end{equation}

In the 5th order equations (\ref{xi-eq}) imply
\begin{equation}\begin{array}{l}\xi_4=\dfrac92w\varkappa_0-4\xi_1-3\xi_2-2\xi_3-\dfrac{\varkappa_0}{2},\\
\xi_5=-\dfrac72w\varkappa_0+3\xi_1+2\xi_2+\xi_3+\dfrac{\varkappa_0}2,\end{array}\end{equation}
from what we have
\begin{equation}\begin{array}{l}\sigma_1=0,\vphantom{\dfrac12}\\
\sigma_8=\dfrac{9w\varkappa_0}{23040\alpha_4}-\dfrac{\alpha_1}{1440\alpha_4}\sigma_2+\dfrac{\alpha_2}{960\alpha_4}\(\sigma_2^2-2\sigma_4\)+ \dfrac{\alpha_3}{720\alpha_4}\(-3\sigma_2^3+18\sigma_4\sigma_2+8\sigma_3^2-24\sigma_6\)+\\ \qquad\quad{}+\dfrac1{48}\sigma_2^4-\dfrac14\sigma_4\sigma_2- \dfrac29\sigma_3^2\sigma_2+ \dfrac23\sigma_6\sigma_2+\dfrac8{15}\sigma_5\sigma_3+\dfrac14\sigma_4^2-\dfrac{\varkappa_0}{23040\alpha_4},\vphantom{\dfrac{\dfrac12}{\dfrac12}}\\
\sigma_{10}=-\dfrac{7w\varkappa_0}{64\cdot8!\alpha_5}+\dfrac{3\alpha_1}{16\cdot8!\alpha_5}\sigma_2-\dfrac{\alpha_2}{4\cdot8!\alpha_5}\(\sigma_2^2-2\sigma_4\)-\dfrac{\alpha_3}{4\cdot8!\alpha_5}\(-3\sigma_2^3+18\sigma_4\sigma_2+8\sigma_3^2-\right.\\ \qquad\quad{}-\left.24\sigma_6\)-\dfrac{105}{8!}\sigma_2^5+\dfrac{2100}{8!}\sigma_4\sigma_2^3+\dfrac{2800}{8!}\sigma_3^2\sigma_2^2-\dfrac{8400}{8!}\sigma_6\sigma_2^2-\dfrac13\sigma_5\sigma_3\sigma_2-\dfrac{6300}{8!}\sigma_4^2\sigma_2+\vphantom{\dfrac{\dfrac12}{\dfrac12}}\\ \qquad\quad{}+\dfrac{25200}{8!}\sigma_8\sigma_2- \dfrac{5600}{8!}\sigma_4\sigma_3^2+\dfrac{19200}{8!}\sigma_7\sigma_3+\dfrac{16800}{8!}\sigma_6\sigma_4+\dfrac15\sigma_5^2+\dfrac{\varkappa_0}{64\cdot8!\alpha_5}.\end{array}\end{equation}

\section{Exponential solution in an arbitrary order (supposition)}

Generalizing these equalities we may suppose that in an arbitrary order equations
\begin{equation}
\sum_{i=1}^p\alpha_i{G^{(i)}}_{\mu\nu}=\varkappa T_{\mu\nu}
\end{equation}
take form
\begin{equation}\begin{array}{l}\left\{\begin{array}{l}\displaystyle\sum^p_{i=1}(2i-1)\xi_i=\varkappa_0,\\
\displaystyle\sum^p_{i=1}\xi_i=w\varkappa_0,\end{array}\right.\end{array}\end{equation}
where
\begin{equation}\begin{array}{l}\xi_i\equiv\alpha_i{G^{(i)j}}_j,\vphantom{\dfrac12}\\
\displaystyle{G^{(i)j}}_j=\dfrac{2^i}{2i-1}\sum_{s_1l_1+\ldots+s_kl_k=2i}(-1)^{l_1+\ldots+l_k+2i-1}\dfrac{(2i)!}{s_1^{l_1}\cdot\ldots\cdot s_k^{l_k}l_1!\cdot\ldots\cdot l_k!}\sigma_{s_1}^{l_1}\cdot\ldots \cdot\sigma_{s_k}^{l_k},\end{array}\end{equation}
so
\begin{equation}\begin{array}{l}\displaystyle\xi_{p-1}=\[\dfrac{2p-1}2w-\dfrac12\]\varkappa_0-\sum_{i=1}^{p-2}(p-i)\xi_i,\\
\displaystyle\xi_p=\[\dfrac{2p-3}{2}w-\dfrac12\]\varkappa_0+\sum_{i=1}^{p-2}(p-i-1)\xi_i,\end{array}\end{equation}
from what we have
\begin{equation}\begin{array}{l}\displaystyle\dfrac{2^{p-1}\alpha_{p-1}}{2p-3}\sum_{s_1l_1+\ldots+s_kl_k=2p-2}(-1)^{l_1+\ldots+l_k+2p-3}\dfrac{(2p-2)!}{s_1^{l_1}\cdot\ldots\cdot s_k^{l_k}l_1!\cdot\ldots\cdot l_k!} \sigma_{s_1}^{l_1}\cdot\ldots \cdot\sigma_{s_k}^{l_k}=\\ \displaystyle\qquad\quad{}=\[\dfrac{2p-1}2w-\dfrac12\]\varkappa_0-\sum_{i=1}^{p-2}(p-i)\dfrac{2^i\alpha_i}{2i-1}\sum_{s_1l_1+\ldots+s_kl_k=2i}(-1)^{l_1+\ldots+l_k+2i-1}\times\\ \displaystyle\qquad\quad{}\times\dfrac{(2i)!}{s_1^{l_1}\cdot\ldots\cdot s_k^{l_k}l_1!\cdot\ldots\cdot l_k!} \sigma_{s_1}^{l_1}\cdot\ldots \cdot\sigma_{s_k}^{l_k},\\
\displaystyle\dfrac{2^p\alpha_p}{2p-1}\sum_{s_1l_1+\ldots+s_kl_k=2p}(-1)^{l_1+\ldots+l_k+2p-1}\dfrac{(2p)!}{s_1^{l_1}\cdot\ldots\cdot s_k^{l_k}l_1!\cdot\ldots\cdot l_k!} \sigma_{s_1}^{l_1}\cdot\ldots \cdot\sigma_{s_k}^{l_k}=\\ \displaystyle\qquad\quad{}=\[\dfrac{2p-3}2w-\dfrac12\]\varkappa_0-\sum_{i=1}^{p-2}(p-i-1)\dfrac{2^i\alpha_i}{2i-1}\sum_{s_1l_1+\ldots+s_kl_k=2i}(-1)^{l_1+\ldots+l_k+2i-1}\times\\ \displaystyle\qquad\quad{}\times\dfrac{(2i)!}{s_1^{l_1}\cdot\ldots\cdot s_k^{l_k}l_1!\cdot\ldots\cdot l_k!} \sigma_{s_1}^{l_1}\cdot\ldots \cdot\sigma_{s_k}^{l_k}.\end{array}\end{equation}
Unfortunately, it is only supposition, but I hope to prove it in the next work.

\section*{Conclusions}
Anisotropic exponential cosmological solutions for a space of arbitrary dimension filled with ordinary matter in the 4th and 5th orders of Lovelock gravity were obtained. Also we have supposed a generalization of such solutions on an arbitrary order.

All the solutions are represented as a set of conditions on Hubble parameters. Unfortunately, it is the problem to write down every Hubble parameter as a fuction of parameters $n$, $\alpha_i$, $w$ and $\varkappa_0$. Moreover, such conditions may be uncompatible. For the second order of Lovelock gravity this problem was investigated in \cite{chirkov-pavluchenko-toporensky}.

\section*{Acknowledgements}

The author is grateful to Alexey V. Toporensky and Alexander A. Reshetnyak for helpful discussions.
The research was fulfilled within the RFBR  Project No. 17-02-01333.


\begin{thebibliography}{99}
\bibitem{lovelock} D. Lovelock. The Einstein tensor and its generalizations // J. Math. Phys., 1971, vol. 12, \symbol{157} 3, pp. 498--501.
\bibitem{kasner} E. Kasner. Geometrical theorems on Einstein cosmologycal equations // Amer J Math, 1921, vol. 43, p. 217.
\bibitem{jacobs}  K. Jacobs. Spatially homogeneous and Euclidean cosmological models with shear // Astrophys. J., 1968, vol. 153, p. 661.
\bibitem{deruelle} N. Deruelle. On the approach to the cosmologycal singularity in quadratic theories of gravity: The Kasner regimes // Nucl. Phys. B, 1989, vol. 327, p. 253--266.
\bibitem{toporensky-tretyakov} A. Toporensky, P. Tretyakov. Power-law anisotropic cosmologycal solution in 5+1 dimensional Gauss-Bonnet gravity // Grav. Cosmol., 2007, vol. 13, pp. 207--210, arXiv:0705.1346v3 [gr-qc].
\bibitem{kirnos-makarenko-pavluchenko-toporensky} I. V. Kirnos, A. N. Makarenko, S. A. Pavluchenko, A. V. Toporensky. The nature of singularity in multidimensional anisotropic Gauss-Bonnet cosmology with a perfect fluid // Gen. Rel. Grav., 2010, vol. 42: p. 2633, arXiv:0906.0140v1 [gr-qc] --- 11 p.
\bibitem{pavluchenko} S. A. Pavluchenko. The general features of Bianchi-I cosmological models in Lovelock gravity // Phys. Rev. D80: 107501, 2009, arXiv:gr-qc/0906.0141v2 --- 9 p.
\bibitem{arbitrary_order} I. V. Kirnos. Some cosmologycal solutions in an arbitrary order of Lovelock gravity // Grav. Cosmol., 2012, vol. 18, No 4, pp. 259--261, arXiv:1501.00107v1 [gr-qc].
\bibitem{kirnos-pavluchenko-toporensky} I. V. Kirnos, S. A. Pavluchenko and A. V. Toporensky. New features of a flat (4 + 1)-dimensional cosmological model with a perfect fluid in Gauss-Bonnet gravity // Grav. Cosmol., 2010, vol. 16, No. 4, pp. 274--282, arXiv:1002.4488v2 [gr-qc].
\bibitem{chirkov-pavluchenko-toporensky} D. Chirkov, S. Pavluchenko and A. Toporensky. Constant volume exponential solutions in Einstein-Gauss-Bonnet flat anisotropic cosmology with a perfect fluid // Gen. Rel. Grav., 2014, vol. 46, p. 1799, arXiv:1403.4625v2 [gr-qc] --- 13 p.

\end{thebibliography}
\end{document}